\UseRawInputEncoding 
\documentclass[twocolumn,aps,floats,floatfix,superscriptaddress]{revtex4}
\usepackage{graphicx,amssymb}
\usepackage{epsfig}
\newcommand{\beq}{\begin{equation}}
\newcommand{\eeq}{\end{equation}}

\newcommand{\br}{{\bf r}}

\newcommand{\bR}{{\bf R}}

\begin{document}
\draft
\title{Metal-insulator transitions in bilayer electron-hole systems in transition metal dicalcogenides}
\author{ S. T. Chui}
\affiliation{ Bartol Research Institute and Department of Physics and Astronomy, 
University of Delaware, Newark, DE 19716, USA}
\author{Ning Wang}
\affiliation{ Department of Physics, Hong Kong University of Science and Technology, Clear Water Bay, Hong Kong}
\author{B. Tanatar} 
\affiliation{ Department of Physics, Bilkent University, 06800 Ankara, Turkey}

\begin{abstract}
We investigated metal-insulator transitions for double layer two-dimensional electron hole systems in 
transition metal dicalcogenides (TMDC) stacked on opposite sides of thin layers of boron nitride (BN). 
The interparticle interaction is calculated by 
including the screening due to the polarization charges at different interfaces, including that at the encapsultion and the substrate of experimental structures.  
We compute and compare the energies of the metallic electron-hole plasma and the newly proposed insulating exciton solid with fixed-node diffusion Monte Carlo simulation including the high valley degeneracy of the electron bands. 
We found that for some examples of current experimental structures, the transition electron/hole density is in an  accessible range of $g\times 10^{12}$\,cm$^{-2}$ with $g$ between 4.1 and 14.5 for spacer thicknesses between 2.5 and 7.5 nm.  Our result raise the possibility of exploiting this effect for logic device applications.

\end{abstract}
\maketitle

There has been much recent interest in bilayer electron-electron (e-e) and electron-hole (e-h)  systems in graphene and transition metal dicalcogenides (TMDC) stacked on opposite sides of a boron nitride (BN) film of thickness $d$\cite{Kim,Li0,Li,Tutuc1,Mak,esolid}.
This follows earlier interests in the physics of the two-dimensional electron gas
in single and double layers in Si-MOSFET and in GaAs heterostructures\cite{Bruna}. An attractive feature of the graphene and the TMDC systems is the high mobility that can be achieved, thus promising practical applications. 

An electron-hole system can exist as a collection of excitons or in a metallic state of an electron-hole plasma, 
as has been observed in bulk Ge under laser excitation\cite{ehp, BRAC,CN}. 
Most previous studies consider the collection of excitons to form a fluid\cite{bec}. 
We recently found that in the parameter range of interest the exciton solid is more stable than the exciton fluid.  This solid exhibits a supersolid behaviour as an approximately quantized Coulomb-drag resistance\cite{esolid}. The metal-insulator transition between the exciton solid and the electron hole plasma is the focus of the present study. 
Metal-Insulator transitions make possible the application of the present structure as logic devices. Graphene has also been considered for this application but because it does not have a band gap, the small but finite current makes the application difficult. For practical applications, the exciton energy should be higher than room temperature. This exciton energy is inversely proportional to the Bohr radius.
The exciton energy is much higher in  the TMDC system\cite{Fogler}, because its Bohr radius, $a_{B}=6$\,\AA\cite{parm}, is much smaller than that in the graphene system,  $a_{B, graphene}=90$\,\AA\cite{pg}, which is comparable to that in GaAs. 
In this paper we include the screening of the Coulomb interaction by the polarization charge density at the interfaces of
typical realistic experimental structures and found that the exciton energy can be an order of magnitude less than previous estimates that does not consider the complete structure .
We
extended our previous fixed node diffusion quamtum  Monte Carlo calculation for the single layer electron gas\cite{chuiimp,cep0} to the current bilayer electron hole system with the screened potential for different spaced thickness $d$. Previous many body calculations\cite{bec,Maezono,Sen,Ho,Swier,Perali,peeters1,cdw} were motivated by the GaAs heterostructures and does not correspond to the parameter range of the TMDC system.  
For bulk Ge, the large number of valleys lower the energy of the electron-hole fluid
at high densities\cite{BRAC}. For the bilayer system there are two hole valleys and 6/12 electron valleys for odd/even number of layers. 
We included this effect in our calculation and found it to be important. For $d$ between 2.5 and 7.5\,nm, we found for typical examples of experimental structures, the metal-insulator 
transition to occur at electron/hole densities in an experimentally accessible range of $g\times 10^{12}$\,cm$^2$ with 
$g$ between 4.1 and 14.5.  We now describe our results in detail.

We consider a typical experimental structure consisting of two TMDC layers  of
thicknesses 2\,nm separated by a boron nitride layer  of thickness $d$ that is of the order of 5\,nm. 
The structure is encapsulated by boron nitride films BN$_{top}$ and BN$_{bottom}$ of thicknesses 10 and 20\,nm, respectively.  On one side of this structure is SiO2; on the other side, vaccum.
Effect of the induced surface polarization (the ``image charge''\cite{Jackson}) for a single two-dimensional
structure  was discussed by Keldysh\cite{Kel} and applied to the
study of excitons in single layer structures with isotropic dielectric constants\cite{Chern}. 
Both BN and the TMDC possess hexagonal symmetry  so that the dielectric constants $\epsilon$  exhibit  z and  xy components. These dielectric constants are given\cite{ebn}
for BN as $\epsilon_{z,BN}=3.76$, $\epsilon_{xy,BN}=6.93$; for WSe$_2$ as $\epsilon_{z,TMDC}=7.6$, 
$\epsilon_{xy,TMDC}=15.7$ and for $SiO_2$, 3.6 . The dielectric properties are  characterized by $\epsilon=(\epsilon_{z}\epsilon_{xy})^{1/2}$ and the ratio $\gamma=\epsilon_{xy}/\epsilon_z$.  The geometric means yield $\epsilon_{BN}=5.1$ and $\epsilon_{TMDC}=10.9$. Furthermore, $\epsilon_{TMDC}/\epsilon_{BN}=2.1$, $\gamma_{BN}=1.35$, and $\gamma_{TMDC}=1.43$.
We have previously considered the case of two thin TMDC films separated by a BN film without considering the effect of the encapsulation\cite{esolid}.
We extended this approach and solved the electrostatics problem of a point charge in our cylindrically symmetric
anisotropic dielectric multilayer structure by separating the space into different regions of different dielectric 
constants and relate the Fourier transform of the potential $\tilde{V}(q)$ in different regions by matching the tangential (normal) components of the electric (displacement) fields at the interfaces. The general solution is then obtained with the transfer matrix method.

\begin{figure}[tbph]
\vspace*{0pt} \centerline{\includegraphics[angle=0,width=8cm]{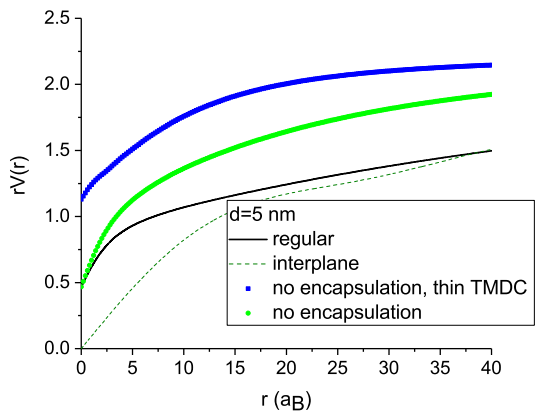}}
\vspace*{0pt}
\caption{The real space intraplane potential $rV_a(r)$ (regular) in units of $e/\epsilon_{BN}$ for $d=5$\,nm  
together with resuts with no encapsulation, no encapsulation and very thin TMDC, and results for the 
interplane potential.}
\label{rvrs}
\end{figure}

From $\tilde{V}(q)$ we have computed the screened potential $V(r)$ in real space. 
As the distance becomes larger than the multilayer thickness the effect of screening diminishes , $rV(r)$ approaches the limit of $ e/<\epsilon>$ with
$<\epsilon>=(1+\epsilon_{SiO2})/2$. 
In Fig.\,\ref{rvrs} we show the intraplane potentials $V_a$ in units of $e/epsilon_{BN}$ close to the origin. 
For no encapsulation and thin TMDCs our previous result\cite{esolid}
shows that for small $r$, 
$rV(r)$ approaches $\epsilon_{BN}/\langle\epsilon'\rangle=1.67$ with
$\langle\epsilon'\rangle=(1+  \epsilon_{BN})/2.$ , 
close to  our numerical result of 1.3 at $r=0$ in this figure.
In the presence of  the TMDC, a very simple effective medium idea suggests that
$\langle\epsilon''\rangle=(\epsilon_{TMDC}+  \epsilon_{BN})/2$.
 Numerically $rV(r)$ approaches 0.5 in this figure, close to our estimate of $\epsilon_{BN}/\langle\epsilon''\rangle=0.64$, 
There is very little
dependence of the intraplane potential on $d$ at small distances. 

\begin{figure}[tbph]
\vspace*{0pt} \centerline{\includegraphics[angle=0,width=8cm]{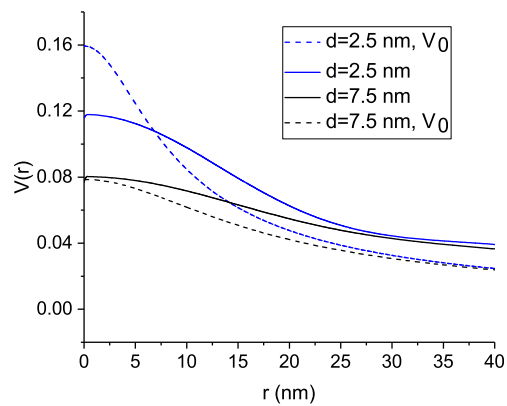}}
\vspace*{0pt}
\caption{The real space interplane potential $V(r)$  for two values of $d$ together with results for the bare Coulomb potential $V_0=e/[d\gamma_{BN}+t_{TMDC}\gamma_{TMDC})^2+r^2]^{1/2}$.}
\label{vr1}
\end{figure}

We next turn our attention to the interplane potential $V_e$ for the interaction
between charges on opposite sides of the BN spacer.
\begin{figure}[tbph]
\vspace*{0pt} \centerline{\includegraphics[angle=0,width=8cm]{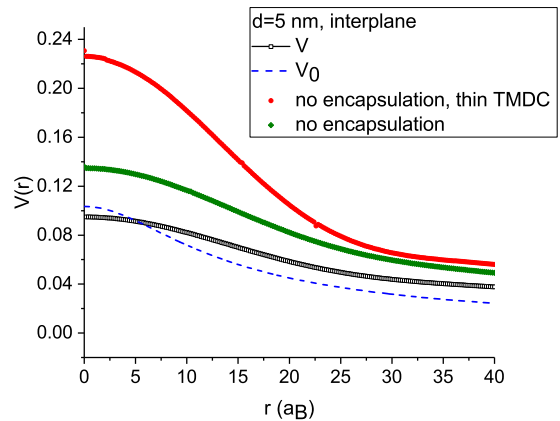}}
\vspace*{0pt}
\caption{The real space interplane potential $V_e(r)$ in units of $e/\epsilon_{BN}/a_{B}$ for d=5nm  together with resuts for the corresponding structures with no encapsulation. no encapsulation and very thin TMDC and the bare Coulomb potential $V_0=e/[d\gamma_{BN}+t_{TMDC}\gamma_{TMDC})^2+r^2]^{1/2}$.}
\label{vrs}
\end{figure}
In Fig.\,\ref{vrs} we show the real space interplane potential $V_e(r)$ in units of $e/\epsilon_{BN}/a_{B}$ for 
$d=5$\,nm  together with results for the corresponding structures with no encapsulation, no encapsulation and 
very thin TMDC, and, for reference, the Coulomb potential in BN 
$V_0=e/[(d\gamma_{BN}+t_{TMDC}\gamma_{TMDC})^2+r^2]^{1/2}/\epsilon_{BN}$ in which $t_{TMDC}$ is the thickness of the TMDC film.
The potential for our structure is softer than $V_0$ because as $r$ increases there is less screening and thus the potential decreases less rapidly. The encapsulation and the finite TMDC thickness produces a significant effect.
for the case without the encapsulation and very thin TMDCs, in our previous study  \cite{esolid} we found that the screened potential for the interaction at small transverse distance $r$ approaches 
$V(r)= V_0(r)\epsilon_{BN}/\epsilon_{eff},$ 
where the effective dielectric constant is given by
$1/\epsilon_{eff}=(1-\beta^2),$
$\beta=(1-\epsilon_{BN})/(\epsilon_{BN}+1)$
is the well known image charge in elementary electrostatics\cite{Jackson}.
Our estimate thus suggests that 
the potential at $r=0$ is approximately 
equal to 0.35.
This is close to the $r=0$ value of 0.39 in this figure.    
In contrast to the intraplane potential, the dependence of the interplane potential on the BN spacer thickness is significant. This is illustrated in Fig.\,\ref{vr1}.

The energy due to the interaction of a charge and the uniform charge backgrounds 
of density $\sigma=1/(\pi r_s^2a_B^2)$ is given by
$$E_{bg}=  [\tilde{V}_e(q=0)-\tilde{V}_a(q=0)]/(\pi r_s^2) Ry.$$ 
If we approximate $V$ by a Coulomb potential with an effective screening constant $E_{bg}= -\sigma  d/(2\epsilon_{eff})$. 
Numerically, we found that 
$-E_{bg} / (\sigma) = 6.19, 10.08, 15 Ry$ for d=2.5 , 5,  7.5 \AA; corresponding to an $\epsilon_{eff}/\epsilon_{BN}=1.1, 0.9, 0.8$. 
For the electrically non-neutral e-e systems, the corresponding background energy is inversely proportional to the first power and not the second power  of $r_s$ and much larger\cite{Bonsall}. The origin of the energy scale is shifted so that most of the energy  per particle are negative whereas  most of the energies  per particle for the present system are positive.

To gain some intuition of the properties of the system, we first discuss the  physical property of a single exciton in the bilayer WSe2 structure with BN in between. 
Because our potential is cylindrically symmetric, the exciton wave function $\chi$ can be separated into a radial 
and an angular component, $\chi(r,\phi)=\psi(r)e^{il\phi}$. 
We discretize the radial equation and solve for the eigenvalue problem of the 
matrix,  which is tridiagonal but not symmetric,   
with the EISPACK routine rt.f in double precision. 

\begin{figure}[tbph]
\vspace*{0pt} \centerline{\includegraphics[angle=0,width=8cm]{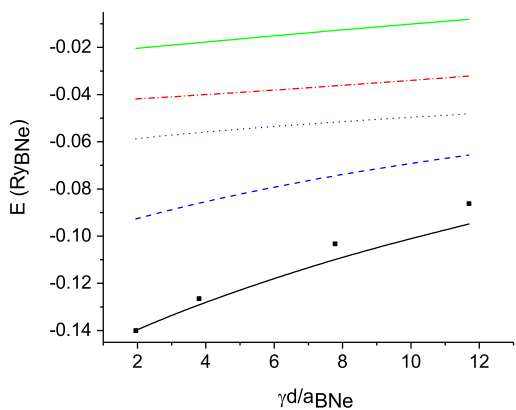}} 
\vspace*{0pt}
\caption{Exciton binding energies in units of Ry as a function of the BN thickness normalized by the exciton Bohr radius for $l=0$. Also shown is an analytic estimate for the ground-state energy (solid black square).}
\label{exce}
\end{figure}

The bound state eigenvalues in units of Ry as a function of differnt BN thicknesses 
are shown in Fig.\,\ref{exce} for $l=0$ (lines) .  Also shown is an analytic
estimate\cite{esolid} for the ground-state energy using the second derivative of the interplane potential, which agrees well the numerical result. Our result is an order of magnitude smaller than previous estimates that uses a potential without the effect of the encapsulation\cite{Fogler}.

\begin{figure}[tbph]
\vspace*{0pt} \centerline{\includegraphics[angle=0,width=8cm]{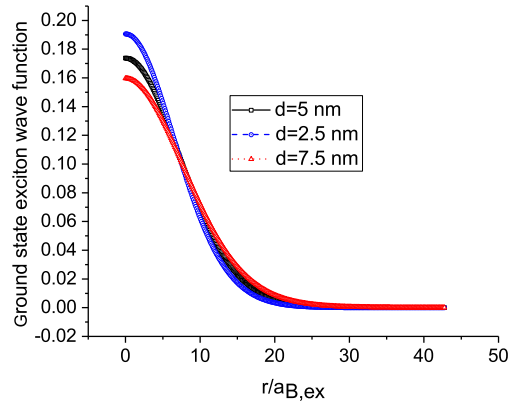}}
\vspace*{0pt}
\caption{ Exciton wave functions for l=0 for different d as a function of radial distance in units of the Bohr radius.}
\label{exgf}
\end{figure}

The ground-state wave functions for three different values of $d$ are shown in Fig.\,\ref{exgf}. 
We have previously considered the large $d$ limit for the potential $V_0$ and found that the width of the 
wave function $\xi$ is of the order of 
$\xi \approx  d(a_B/d)^{1/4}$. 
The size of the wavefunction is larger than this estimate because our potential is softer. Also, because the 
potential does not change much as $d$ is decreased, the ground-state
wavefunction has a weak dependence on $d$. The electroluminescence is proportional to the probability of 
finding the electron and hole on top of each other and thus equal to $\psi_{exciton}(r=0)^2$.  We next turn our attention to the metal insulator transition. (MIT)

To investigate the MIT, we have calculated the energies of the electron-hole plasma and 
the hexagonal quantum exciton solid  by extending our previous 
fixed-node diffusion Monte Carlo simulation\cite{cep0,bt,chuiimp}. 
The calculation starts off with a variational trial wavefunction of the form\cite{bt}
\beq
\Psi=D\exp\sum_{i<j}u(r_{ij})
\label{vwf}
\eeq
that is a product of a Slater determinant $D$ and a Jastrow correlation factor. We next discuss
the choice of $u$.

Our interparticle interaction is given by $\sum_{ij}V_{ij}\rho_i(q)\rho_j(-q)/2$,
where $V_{ii}=V_a$ ($V_{i\neq j}=V_e$) is the intraplane (interplane) interaction.
For the electron-electron fluid system, the simplest way  is just to  have
\beq
u_{ij}=-1/S+[1/S^2+V_{ij}]^{1/2}
\label{lu1}
\eeq
 where $S$ is the static structure factor for noninteracting electrons.
For the electron-hole system, $u_{12}$ becomes imaginary for $V_e<0$. Thus, this approach cannot be used. De Palo  {\it et al}.\cite{Sen} have investigated the phase diagram of  a  two spin 2 layer particle hole system  interacting with an isotropic Coulomb potential 
with fixed node Monte Carlo simulation in a different parameter regime of $d/a_B< 3.$  We have used a modification of their u.

We introduce new operators that are linear combinations of charge densities $\rho_{1,2}$  on the two layers:
\beq
\rho_{\pm}=(\rho_1\pm\rho_2)/r
\label{css}
\eeq
with the normalization factor $r=2$ as $\rho$ involves the product of two field operators.
If we interpret 1 as pseudospin up and 2 as pseudospin down, then $\rho_+$ 
($\rho_-$) is t
he particle (pseudospin) density.
The interparticle interaction  can be written in diagonal form 
as $|\rho_+|^2 V_++|\rho_-|^2 V_-.$
$V_{\pm}=(V_a\pm V_e)/2.$ 
This motivated a trial wave function proportional to $\exp{(u_+\rho_+^2+u_-\rho_-^2)}$
where 
\beq
2u_{\pm}=-1/S+[1/S^2+4mV_{\pm}/(\hbar^2k^2]^{1/2}.
\label{lu2}
\eeq
The exponent becomes $(u_++u_-)(\rho_1^2+\rho_2^2)+2(u_+-u_-)\rho_1\rho_2.$
Rapisarda\cite{Sen} and coworkers used a correlation factor 
\beq
4u'_{\pm}=-1/S+[1/S^2+8mV_{\pm}/(\hbar^2k^2]^{1/2}.
\label{lu3}
\eeq
This can be interpreted as a different normalization factor $r=\sqrt{2}$.
For the electron-electron system, in the limit that $d$ approaches zero so that $V_a=V_e$ and $V_-=0$, we get back the correct limit of correlation involving the particle-particle correlation. This is not true with the correlation factor in Eq. \ref{lu3}. 
We found numerically that Eq.\,(\ref{lu1}) gives the lowest variational energy where possible.
Eq.\ref{lu3} gives the highest energy.
A similar generalization is used for the solid case.

\begin{figure}[tbph]
\vspace*{0pt} \centerline{\includegraphics[angle=0,width=8cm]{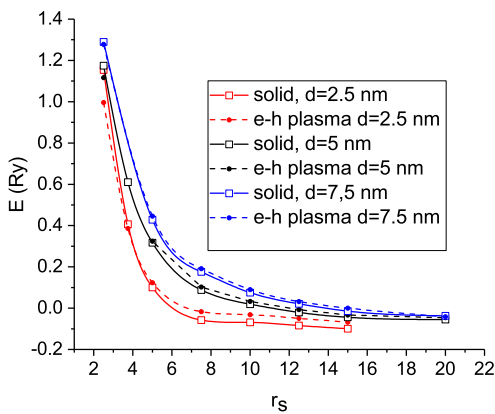}}
\vspace*{0pt}
\caption{The energies in units of Ry$_{BN}$ of the solid (solid lines) and the e-h plasma (dashed lines) as a function of $r_s$ from fixed-node MC simulations 
for three values of $ d_{BN}$.}
\label{fse}
\end{figure}
We have performed calculations with a sample of 30 electrons and 30 holes
for different number of electron valleys $n_v$(1,2,3,5) and doubly degenerate holes.
The energies are then fitted with a quadratic polynomial in $1/\sqrt{n_v}$ and extrapolated to provide for estimates of the energies for $n_v=6,\ 12.$
Our results are expressed in terms of the usual dimensionless density parameter $r_s$ defined by 
$\pi (r_sa_{B})^2=1/\sigma.$ 
The energies for $n_v=6$ in units of $Ry$ of different phases as a function of $r_s$ from fixed-node diffusion MC simulations for 
three values of $d$  are shown in Fig.\,\ref{fse}. 
At small $r_s$, we expect the kinetic energy term that is proportional to $1/r_s^2$ to dominate. The intraplane 
potential energy that is proportional to $-1/r_s$ comes in as $r_s$ increases. 
For large $r_s$ the solid energy per particle is of the order of half the exciton energies. This is consistent with results in Fig. \ref{exce}. 
Our energy may be measured experimentally as it is equal to  $e^2\sigma /2/(C/A)$ where $C/A$ is the total capacitance (C) per unit area (A) of the system. The total capacitance contains contributions from the self  capacitances due to the e-e and h-h interactions and the mutual capacitance from the e-h interaction.
Such type of capacitance measurement have recently been carried out by Ma and corworkers\cite{Cornell} on similar structures.

\begin{figure}[tbph]
\vspace*{0pt} \centerline{\includegraphics[angle=0,width=8cm]{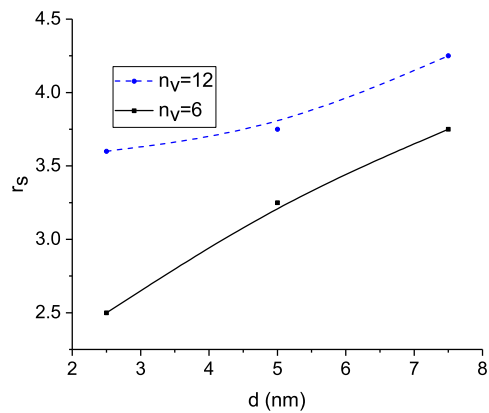}}
\vspace*{0pt}
\caption{ The transition $r_s$ as a function of $ d_{BN}$. }
\label{rsc}
\end{figure}

From where the solid lines cross the dashed lines we obtain the phase boundary of the transition in Fig.\,\ref{rsc}. The transition electron/hole density is in an experimentally accessible range of $g\times 10^{12}$\,cm$^{-2}$ with $g$ between 4.1 and 14.5.

In conclusion, we have investigated the metal-insulator transition between the electron hole plasma and a newly proposed exciton solid for double layer two-dimensional 
electron-hole systems  in transition metal dicalcogenides stacked on opposite sides of thin layers of BN. We computed the energies of the two phases with fixed-node diffusion  Monte Carlo simulation. The valley degeneracy of the bands lowers the energy of the e-h plasma by a significant amount. 
The screened interlayer and intralayer Coulomb interaction is calculated for typical realistic experimental structures by including the  
polarization charges at the interfaces of encapsultions and substrates. We found that the encapsulation can lower the exciton energy by an order of magnitude. For spacer thickness $d$ between 2.5 and 7.5\,nm, the transition electron/hole density is in an experimentally accessible range of $g\times 10^{12}$\,cm$^{-2}$ with $g$ between 4.1 and 14.5. 
We have explored the effect of the potential from the charges on the boron and the nitrogen ions and found that it producd a less than ten per cent change in the energy difference between the two phases\cite{unp}. The phase boundary is not affected. 
 Our results raise the possibility of exploiting this effect for logic device applications.

In this paper we focus on the case with equal number of electrons and holes. In the metallic phase, when the number of electrons is not equal to the number of holes, the state will remain metallic.  In the insulating phase, we  expect a small number of the excess particles/holes to form a Wigner solid occupying the interstitual positions of the exciton solid, since the Wigner solid is stable at low densities.   This Wigner solid is further stabilized by an enahnced effective mass for the excess particles to hop between interstitual positions of the exciton solid and by the their interation with the Boron and the nitrogen ions. The system remains insulating. Eventually when the imbalance gets bigger the Wigner solid becomes unstable and the system becomes metallic. This metal insulator transition provides for another possibility to making logic devices.

\acknowledgements{N. Wang thanks the support from the 
 National Key R\&D Program of China (2020YFA 0309600/0309602) and the Research Grants Council of Hong Kong (Project No. 16303720) }.

\end{document}